\documentclass[useAMS,usenatbib]{mn2e}

\usepackage{aas_macros}
\usepackage{amsmath}
\usepackage{graphicx}
\usepackage{epstopdf}
\usepackage{float}
\usepackage{natbib}
\usepackage{wasysym}

\title[Viscous time lags between starburst and AGN activity]{Viscous time lags between starburst and AGN activity}
\author[M. Blank and W. J. Duschl]{Marvin Blank$^{1}$\thanks{E-mail:
mblank@astrophysik.uni-kiel.de (MB); wjd@astrophysik.uni-kiel.de (WJD)} and Wolfgang J.
Duschl$^{1,2}$\footnotemark[1]\\
$^{1}$Institut f\"{u}r Theoretische Physik und Astrophysik,
      Christian-Albrechts-Universit\"{a}t zu Kiel, Leibnizstr. 15, D-24118 Kiel, Germany\\
$^{2}$Steward Observatory, The University of Arizona, 933 N. Cherry Ave., Tucson, AZ 85721, USA}

\begin{document}

\date{\today}

\pagerange{\pageref{firstpage}--\pageref{lastpage}} \pubyear{2016}

\maketitle

\label{firstpage}

\begin{abstract}
There is strong observational evidence indicating a time lag of order of some 100\,Myr
between the onset of starburst and AGN activity in galaxies.
Dynamical time lags have been invoked to explain this.
We extend this approach by introducing a viscous time lag the gas
additionally needs to flow through the AGN's accretion disc
before it reaches the central black hole.
Our calculations reproduce the observed time lags and are in accordance
with the observed correlation between black hole mass and stellar velocity dispersion.
\end{abstract}


\section{Introduction}
Motivated by, for instance, observed correlations between the mass of an AGN's central black
hole and the host galaxy's velocity dispersion \citep[e.g.,][]{2000_Gebhardt_Bender_Bower} and between black
hole mass and bulge mass \citep[e.g.,][]{1995ARA&A..33..581K}, there is an ongoing debate whether,
and if so, how starbursts and AGN are connected to each other.

\citet{2005_Di-Matteo_Springel_Hernquist}, for instance, explain such correlations as due to a thermal AGN feedback
that heats the gas of the galaxy and thus prevents further star formation and AGN activity:
More massive galaxies have a deeper gravitational potential well, thus the black hole has to gain more mass
before its luminosity is capable of expelling the gas from the galaxy and quenching star formation and AGN activity.
This then leads to the velocity dispersion and the bulge mass, resp., to be related to the black hole mass. 
In these simulations starburst and AGN activity occur simultaneously, but recent observations show that AGN
activity may be delayed with regard to star formation activity by time-scales of 50--250\,Myr
\citep[e.g.,][]{2007_Davies_Mueller-Sanchez_Genzel, 2009ApJ...692L..19S, 2010MNRAS.405..933W}.

\citet{2012MNRAS.420L...8H} argues that such a time lag can occur for purely dynamical reasons.
His high spatial resolution simulations of galaxy mergers show first an inward motion of gas towards
the dynamical centre giving rise to (a burst of) star formation.
In these models, the gas flowing further inwards can do so only by losing angular momentum by
gravitational instabilities.
This, in turn, gives rise to a time lag between star formation and AGN activity.

We extend this idea by modelling the loss of angular momentum and the ensuing inflow in the
framework of an accretion disc scenario.
Thus the time lag between starburst and AGN activity consists of a dynamical lag given by the time span the gas needs
to reach the accretion disc and a subsequent viscous lag given by the time span the gas needs to flow through
the accretion disc until it reaches the black hole.

In Section \ref{sec:numerics} we explain our numerical methods and the setup of the merger event that is,
in this scenario, responsible for the inflow of gas to the newly forming galactic centre.
In Section \ref{sec:results} we present and discuss the general picture that results from our calculations.
As our model depends on a number of parameters, we perform a parameter study in Section \ref{sec:param} to
show the robustness of our results against parameter changes.
In Section \ref{sec:summary} we summarise our findings.

\section{Numerical Methods}\label{sec:numerics}
We simulate galaxy collisions using the TreeSPH code {\sc gadget-2} \citep{2005MNRAS.364.1105S}.
Radiative cooling of an optically thin primordial gas in ionization equilibrium is taken into
account following \citet{1996ApJS..105...19K}.
Additionally we include star formation, AGN evolution and AGN feedback as described in the following subsections.

\subsection{Star formation}\label{sec:sf}
Following \citet{2005MNRAS.364..552S} a gas particle is considered for star formation if its density
$\rho$ exceeds a critical density $\rho_{\rmn{crit}}$ and it is in a converging flow ($\rmn{div}\,v < 0$).
Then a star particle is created with a probability
\begin{equation}
 p = \frac{m}{m_{*}} \left[ 1 - \exp \left( - \frac{c_{\rmn{eff}} \Delta t}{t_{\rmn{g}}} \right)\right]
 \label{eq:prob}
\end{equation}
\citep[see][]{2003MNRAS.339..289S}
where $m$ is the gas particle's mass, $m_{*} = m_{0}/N_{\rmn{g}}$ the mass of the star to be formed,
$m_{0}$ the initial gas particle mass, $N_{\rmn{g}}$ the number of stars that can be formed from one gas particle,
$c_{\rmn{eff}}$ the star forming efficiency, $\Delta t$ the respective time step of the code and
$t_{\rmn{g}}=\sqrt{3 \rmn{\pi} / 32 G \rho}$ the free-fall time of the gas particle.
If $m < 1.5 \times m_{*}$, condition (\ref{eq:prob}) is dropped and the gas particle is directly converted into a star particle.
Gas particles can only reduce their mass due to star formation.
In our simulations star particles are collisionless particles and interact with other particles only via gravitational forces.
For the star forming efficiency we use a value of $c_{\rmn{eff}}=0.1$ following \citet{1992ApJ...391..502K}.
The parameter $N_{\rmn{g}}$ determines the numerical resolution of the star formation rate,
like \citet{2005MNRAS.364..552S} we use a value of $N_{\rmn{g}} = 2$.
We furthermore set $\rho_{\rmn{crit}} = 4.5 \times 10^{8}~\rmn{M}_{\rmn{\astrosun}}\,\rmn{kpc}^{-3}$.

With these parameters the simulation of an isolated galaxy characterized by the parameters given in
Table~\ref{tab:parameters} gives an average star formation rate of $0.26\,\rmn{M}_{\rmn{\astrosun}}\,\rmn{yr}^{-1}$.
This value fits well with the star formation rate that is expected in such a galaxy:
According to \citet{1998_Kennicutt} the star formation rate of an isolated galaxy can be estimated as
\begin{equation}
 \dot{M}_{\rmn{SFR}} = 0.017 \frac{M_{\rmn{gas}}}{\tau_{\rmn{dyn}}}
 \label{eq:sfr}
\end{equation}
with the total gas mass $M_{\rmn{gas}}$ and the dynamical time-scale at the half gas-mass radius $\tau_{\rmn{dyn}}$.
According to this equation an isolated galaxy characterized by the parameters given in
Table~\ref{tab:parameters} has a star formation rate of $0.3\,\rmn{M}_{\rmn{\astrosun}}\,\rmn{yr}^{-1}$.


In our simulations star formation takes place only in the host galaxy and not in the accretion disc
that surrounds the central black hole.
It is currently highly debated if such accretion discs can be subject to star formation.
On the one hand, the accretion disc's viscosity might heat the gas up to temperatures that 
prevent the gravitational collapse of the gas and thus subsequent star formation.
On the other hand gravitational fragmentation of a self-gravitating accretion disc might
favor gravitational collapse and thus trigger star formation.   
Star formation in the accretion disc will subsequently lead to stellar feedback processes,
which can potentially clear the central region of gas and thus prevent accretion of material towards the black hole.
However, as this topic is still a subject of intense debate
we will neglect star formation in the accretion disc in our simulations.
Stellar feedback is not included in our simulations, we
leave the study of these effects for later investigations.

\subsection{Modeling the AGN}
The AGN is represented by an \textit{accretion disc particle} (ADP) as introduced by \cite{2011MNRAS.412..269P},
which interacts with its environment only via gravitational forces, accretion of gas particles and AGN feedback.

The black hole growth rate is calculated via a subgrid model: The ADP contains a black hole and an accretion disc,
the mass accreted by the ADP is added to the outer rim of the accretion disc, from where it is accreted
towards the black hole. The details of this accretion process are described in the next subsection.

We use the implementation of \citet{2005A&A...435..611J} to describe the accretion of gas particles, but
as suggested in \cite{2011MNRAS.412..269P} gas particles are accreted only if their distance to the ADP
falls below the ADP's accretion radius $R_{\rmn{acc}}$, which is a free parameter.

\citet{2013MNRAS.431..539W} investigate the impact of various parameters in the ADP model in
simulations of Milky Way-sized galaxy mergers and explore how to set $R_{\rmn{acc}}$ accordingly.
Their simulations produce unphysical results when $R_{\rmn{acc}} \le 0.02\,h_{\rmn{min}}$ or
$R_{\rmn{acc}} \ge 0.2\,h_{\rmn{min}}$, where $h_{\rmn{min}}$ is the minimum smoothing length.
Their definition of physical or unphysical is based on, besides analyzing the structural evolution
of the galaxy merger, how close the merger remnant lies on the $M_{\rmn{BH}}$-$\sigma$ correlation.
For very small values of $R_{\rmn{acc}}$ the accretion rate of the black hole is underestimated
which leads to black hole masses that are too small,
whereas very large values of $R_{\rmn{acc}}$ lead to black hole masses that are too large.

In our simulations the minimum smoothing length is $h_{\rmn{min}} = 250\,\rmn{pc}$ and thus
in our reference model $R_{\rmn{acc}} = 0.8\,h_{\rmn{min}}$.
This is much larger than allowed by the analysis of \citet{2013MNRAS.431..539W}.
However, our simulations include AGN feedback in the form of a wind that reduces
the accretion rate of the black hole and thus the final black hole mass.
Therefore our simulations produce black hole masses that are in agreement with the observed $M_{\rmn{BH}}$-$\sigma$ correlation,
making an accretion radius of $R_{\rmn{acc}}=200\,\rmn{pc}$ for our reference model a reasonable choice.


\subsection{Evolution of the AGN}\label{sec:AGNevol}
We assume the accretion disc that is hosted by the ADP to be rotationally symmetric and geometrically thin,
furthermore we assume the gravitational potential to change slowly enough so that in the following we can neglect
its explicit time derivative.
This then leads to a single equation that describes the time dependent evolution of the accretion disc's
surface density $\Sigma$ \citep[see, e.g.,][]{1981ARA&A..19..137P}:
\begin{equation}
 \frac{\upartial \Sigma}{\upartial t} + \frac{1}{s} \frac{\upartial}{\upartial s} \left[ \frac{\frac{\upartial}{\upartial s} \left( \nu \Sigma s^3 \frac{\upartial \omega}{\upartial s} \right)}{\frac{\upartial}{\upartial s} \left( s^2 \omega\right)} \right] = 0
 \label{eq:evolution}
\end{equation}
Here $s$ is the distance to the central black hole. $\omega$ is the angular frequency which we calculate
through a balance between gravitational and centrifugal forces.
$\nu$ is the viscosity of the gas, which we parametrize such that the viscous time-scale is related to the dynamical
time-scale via $t_{\rmn{visc}} = \xi t_{\rmn{dyn}}$ with a constant parameter $\xi$ of order $10^2$--$10^3$
\citep{2000_Duschl_Strittmatter_Biermann}\footnote{\citet{2000_Duschl_Strittmatter_Biermann} use the parameter $\beta = \xi^{-1}$,
which is of order $10^{-3}$--$10^{-2}$.}.

We carry out the time integration of equation~(\ref{eq:evolution}) by applying an implicit Crank-Nicolson finite
differences scheme \citep{1947_Crank_Nicolson_Hartree}.
The radial calculation domain extends from an inner radius $s_{\rmn{in}}$ to the
disc's initial outer radius $s_{\rmn{out,0}}$.
We set $s_{\rmn{in}}=10^{-3}\,\rmn{pc}$ for all calculations presented in this paper, 
tests showed that moderate variations of this value have almost no effect on the results.
Between the inner and the outer radius 25 grid points are distributed logarithmically.

For solving equation~(\ref{eq:evolution}) with the aim of obtaining the mass supply rate $\dot{M}_{\rmn{d}}$ from
the disc, an initial condition for the surface density is necessary, for simplicity we assume
a constant density distribution $\Sigma_{0} = M_{\rmn{d,0}}/(\rmn{\pi} s_{\rmn{out,0}}^2)$,
with the initial disc mass $M_{\rmn{d,0}}$ and the disc's initial outer radius $s_{\rmn{out,0}}$.
In addition to being a simple initial condition, it guarantees that in the disc viscous processes dominate the time-scales
as initially most of the disc's mass is at larger radii.

Thus for our disc model we have to specify four parameters: the ratio of viscous time-scale to dynamical time-scale $\xi$,
the initial black hole mass $M_{\rmn{BH,0}}$, the initial disc mass $M_{\rmn{d,0}}$ and the disc's
initial outer radius $s_{\rmn{out,0}}$ with the latter corresponding to the ADP's accretion radius $R_{\rmn{acc}}$.
We use $\xi=250$, $M_{\rmn{BH,0}}=5 \times 10^{5}\,\rmn{M}_{\rmn{\astrosun}}$ and
$M_{\rmn{d,0}}=5 \times 10^{5}\,\rmn{M}_{\rmn{\astrosun}}$.
The initial masses of the black hole and the accretion disc have almost no effect on
the results as long as they are not too large.
We furthermore set $R_{\rmn{acc}} = 200\,\rmn{pc}$ where viscous processes start to play a dominant role
for the accretion of material onto the black hole.

\subsection{AGN feedback}
For modeling the AGN feedback we follow \cite{2011_Debuhr_Quataert_Ma,2012_Debuhr_Quataert_Ma}.
There the AGN's radiation causes a force
\begin{equation}
 \dot{p}_{\rmn{rad}} = \tau \frac{L}{c}
\end{equation}
on the surrounding gas where $L = \eta c^2 \dot{M}_{\rmn{BH}}$ is the AGN's luminosity, $\dot{M}_{\rmn{BH}}$
the black hole accretion rate and $\eta$
the accretion efficiency, which is $\sim\,10^{-1}$ for standard accretion discs \citep{1973A&A....24..337S}.

$\tau$ is the wavelength-averaged optical depth of the gas that surrounds the AGN and a free parameter,
following \cite{2011_Debuhr_Quataert_Ma} we use a value of $\tau = 25$.

The AGN additionally drives a wind adding a force
\begin{equation}
 \dot{p}_{\rmn{w}} = \tau_{\rmn{w}} \frac{L}{c}
 \label{eq:force_wind}
\end{equation}
to the surrounding gas where $\tau_{\rmn{w}}$ is a further free parameter of the model determining the
total momentum flux in the wind, we choose a value of $\tau_{\rmn{w}} = 3$
to normalize the $M_{\rmn{BH}}$-$\sigma$ correlation (see Section \ref{sec:results}).

Assuming the wind to be launched at a speed $v_{\rmn{w}}$, equation~(\ref{eq:force_wind}) yields its mass flow
\begin{equation}
 \dot{M}_{\rmn{w}} = \tau_{\rmn{w}} \frac{L}{c v_{\rmn{w}}} = \tau_{\rmn{w}} \eta \frac{c}{v_{\rmn{w}}} \dot{M}_{\rmn{BH}} \, .
 \label{eq:mflow_wind}
\end{equation}
Here we use $v_{\rmn{w}} = 10^4\,\rmn{km}\,\rmn{s}^{-1}$ following \cite{2012_Debuhr_Quataert_Ma}.

As thus not all of the mass that is made available by the accretion disc enters the black hole,
its accretion rate must be reduced accordingly
\begin{equation}
 \dot{M}_{\rmn{BH}} = \dot{M}_{\rmn{d}} - \dot{M}_{\rmn{w}}
\end{equation}
which yields, considering equation~(\ref{eq:mflow_wind}),
\begin{equation}
 \dot{M}_{\rmn{BH}} = \frac{\dot{M}_{\rmn{d}}}{1 + \tau_{\rmn{w}} \eta c v_{\rmn{w}}^{-1}} \,.
 \label{eq:bhar}
\end{equation}

Following, e.g., \cite{2011_Debuhr_Quataert_Ma,2012_Debuhr_Quataert_Ma}, \citet{2005_Di-Matteo_Springel_Hernquist}, \citet{2005MNRAS.361..776S}
we furthermore limit the black hole accretion rate to the Eddington limit 
\begin{equation}
 \dot{M}_{\rmn{Edd}} = \frac{M_{\rmn{BH}}}{\tau_{\rmn{S}}}
 \label{eq:edd}
\end{equation}
\citep{1921_Eddington}.
Here $M_{\rmn{BH}}$ is the black hole mass and $\tau_{\rmn{S}} = \eta \times 4.5 \times 10^{8}\,\rmn{yr}$
the Salpeter time-scale \citep{1964ApJ...140..796S}.

Each time step the total amount of mass $\dot{M}_{\rmn{w}} \Delta t$ and momentum $(\dot{p}_{\rmn{rad}}+\dot{p}_{\rmn{w}}) \Delta t$
made available by the AGN are calculated and equally distributed among the SPH-particles within a radius $2 R_{\rmn{acc}}$.

\subsection{Initial conditions and parameters}
We model our galaxies as described by \citet{1999MNRAS.307..162S} and \citet{2000MNRAS.312..859S} based on the analytical
model of \citet{1998MNRAS.295..319M}.

\begin{table}
  \caption{Parameters of our reference model, a galaxy with these parameters has a scale radius of 3.8 kpc.}
  \label{tab:parameters}
  \begin{tabular}{lll}
    \hline
      virial velocity                       &   $v_{\rmn{200}}$   &   $160\,\rmn{km}\,\rmn{s}^{-1}$   \\
      mass fraction of the disc             &   $m_{\rmn{d}}$     &   $0.041$                         \\
      mass fraction of the bulge            &   $m_{\rmn{b}}$     &   $0.0137$                        \\
      mass fraction of the gas              &   $f_{\rmn{g}}$     &   $0.3$                           \\
      scale height of the disc$^\rmn{a}$    &   $f_{\rmn{d}}$     &   $0.2$                           \\
      scale radius of the bulge$^\rmn{a}$   &   $f_{\rmn{b}}$     &   $0.1$                           \\
      halo concentration                    &   $c_{\rmn{halo}}$  &   $15$                            \\
      spin parameter of the halo            &   $\lambda$         &   $0.05$                          \\
      angular momentum fraction of the disc &   $j_{\rmn{d}}$     &   $0.041$                         \\
    \hline
  \end{tabular}
  $^\rmn{a}$ in fractions of the disc's scale radius
\end{table}

\begin{figure*}
  \includegraphics[width=177mm]{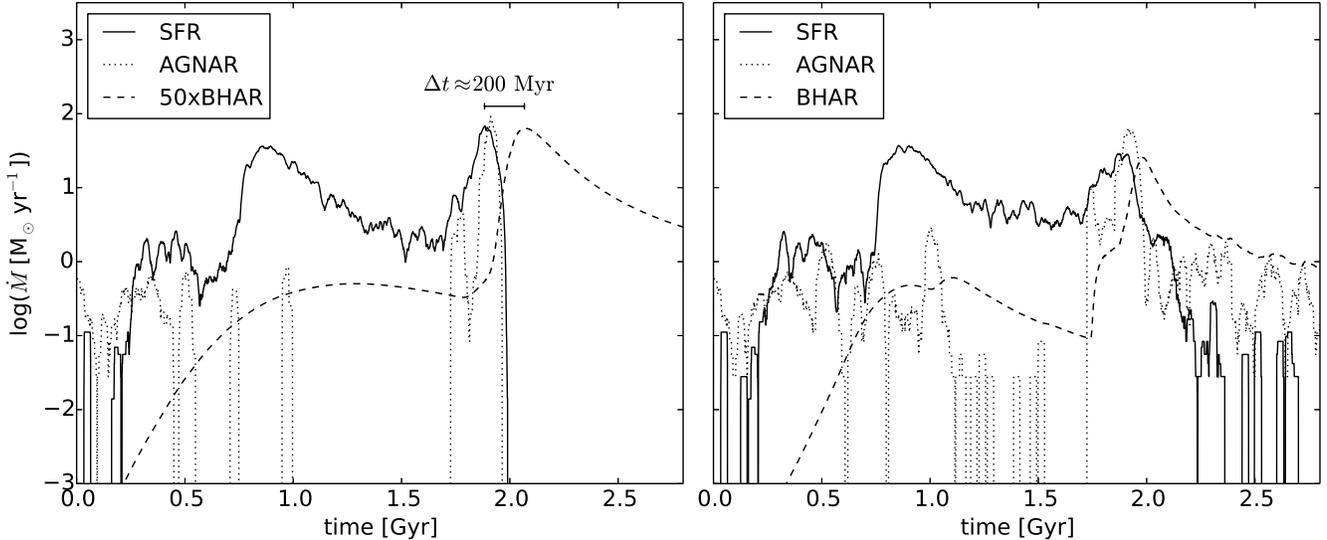}
  \caption{Star formation rate (SFR), accretion rate of the AGN (AGNAR) 
           and black hole accretion rate (BHAR) as functions of time.
           Left: with AGN feedback, right: without AGN feedback.}
  \label{fig:massflows}
\end{figure*}

A galaxy is characterized via a virial velocity $v_{\rmn{200}}$ that corresponds to a virial mass
\begin{equation}
 M_{\rmn{200}} = \frac{v_{\rmn{200}}^3}{10 G H_0}
 \label{eq:init_mass}
\end{equation}
with the gravitational constant $G$ and the Hubble constant\footnote{We
use a value of $H_0 = 70\,\rmn{km}\,\rmn{s}^{-1}\,\rmn{Mpc}^{-1}$ throughout this paper.} $H_0$.

A fraction $m_{\rmn{d}}$ of the virial mass forms a thin (galactic) disc with a profile
\begin{equation}
 \rho_{\rmn{d}}(R,z) = \rho_{0} \exp \left( - \frac{R}{R_{\rmn{d}}} \right) \rmn{sech}^2 \left( \frac{z}{z_{0}} \right) \,.
\end{equation}
The disc's scale height $z_{0}$ is a fraction $f_{\rmn{d}}$ of the disc's scale radius $R_{\rmn{d}}$.
A fraction $f_{\rmn{g}}$ of the disc consists of gas, the rest consists of old stars.
Another fraction $m_{\rmn{b}}$ of the virial mass forms a spherical bulge with a \citet{1993_Hernquist}
profile whose scale radius is again a fraction $f_{\rmn{b}}$ of the disc's scale radius $R_{\rmn{d}}$.

The rest of the total mass forms a dark matter halo with a \citet{1996ApJ...462..563N, 1997ApJ...490..493N} profile
whose scale radius is a fraction $c_{\rmn{halo}}^{-1}$ of the virial radius where $c_{\rmn{halo}}$ is the halo concentration.
The angular momentum of the halo is characterized by a dimensionless spin parameter $\lambda$.
Setting the disc's angular momentum to a fraction $j_{\rmn{d}}$ of the
halo's angular momentum determines the disc's scale radius.
Thus the galaxy model depends on nine parameters,
which we list in Table~\ref{tab:parameters} for our reference model.
The particle's positions are initialized according to the above mentioned mass distributions, the particle's
velocities are initialized according to the solution of the Boltzmann equation. For details on this procedure
we refer to \citet{1999MNRAS.307..162S}.

Each galaxy consists of 30\,000 particles for the halo, 10\,000 for the bulge,
20\,000 for the stellar disc and 20\,000 for the gaseous disc.

According to \citet{2003PhDT.........1K} merging dark matter halos are mostly on parabolic orbits,
we therefore set the galaxies on a parabolic orbit with a periapsis of $r_{\rmn{p}}=5~\rmn{kpc}$ and
an initial distance of $r_{\rmn{0}}=300~\rmn{kpc}$.
In one of the galaxy's centres we place an ADP with mass $10^{6}~\rmn{M}_{\rmn{\astrosun}}$ (sum of initial
black hole mass and initial disc mass).

Test calculations showed that simulations where both galaxies contain an ADP
produce about the same results. We prefer simulations with only one galaxy hosting
an ADP because otherwise we would have to introduce 
a number of additional assumptions and parameters
to account for the merging of the two ADP.

The gravitational softening length for the halo particles is 0.8 kpc and 0.5 kpc for all other particles.
The minimum smoothing length is set to half the gravitational softening length.

\section{The general picture}\label{sec:results}
The time evolution of the merger event brings the galaxies to their first passage $\sim 0.7$\,Gyr after
the start of the simulation, they finally merge at 1.8\,Gyr and form a gas-poor elliptical galaxy.
Fig.~\ref{fig:massflows} shows the star formation rate (SFR), the accretion rate of the AGN (AGNAR) and
the black hole accretion rate (BHAR) as functions of time.
We simulated the merger event once with and once without AGN feedback.

The merging event drives huge amounts of gas towards the centre of the new forming merged galaxy
where it causes a starburst with a lifetime of 200--300\,Myr (in accordance with the observations
of \citealt{2010MNRAS.405..933W} and the simulations of \citealt{2005_Di-Matteo_Springel_Hernquist}) and feeds the AGN.

After being devoured by the ADP the gas has to flow through the accretion disc on a viscous time-scale
that corresponds to a time lag of approx. 200\,Myr between the peak of the AGNAR
and the peak of the BHAR, resulting in a time lag between starburst and AGN activity that is in agreement
with observations \citep[e.g.,][]{2010MNRAS.405..933W}.

This time lag does not render AGN feedback ineffective.
Although the BHAR peaks much later than the SFR and the AGNAR, the feedback already starts working well before the BHAR peaks.
This is evident when comparing the simulations with and without feedback in Fig.~\ref{fig:massflows},
where AGN feedback clearly quenches the SFR and the AGNAR.

However, during the time lag until the onset of AGN activity huge amounts of mass are inserted into the accretion disc.
Thus the BHAR is not quenched by the AGN feedback, the AGN continues evolving for some time after the galaxies have merged.
But due to the feedback the BHAR is reduced by approx. one order of magnitude, compared with the simulation without feedback.
Feedback given to the accretions disc's gas is parametrized by the reduction of the black hole accretion rate
according to equation~(\ref{eq:bhar}).

To check the agreement of our model with the observed correlation between black hole mass and stellar velocity
dispersion ($M_{\rmn{BH}}$-$\sigma$ correlation) we repeat the calculations for different virial masses.
The critical density for the onset of star formation is rescaled according to
\begin{equation}
 \rho_{\rmn{crit,*}} = \frac{M_{\rmn{200,*}}}{M_{\rmn{200,ref}}} \rho_{\rmn{crit,ref}}
\end{equation}
where $\rho_{\rmn{crit,*}}$ and $M_{\rmn{200,*}}$ are the critical density and the virial mass of the
respective model and $\rho_{\rmn{crit,ref}}$ and $M_{\rmn{200,ref}}$ are the critical density and the virial mass of the
reference model that we have specified in the previous sections.
The rescaling of the critical density is necessary to compensate for the changing resolution that
arises from a constant particle number and a changing galaxy mass.
Tests that were conducted in the same manner than described in Section \ref{sec:sf}
show that the simulated star formation rate and the star formation rate implied by equation~(\ref{eq:sfr}) differ
by less than a factor of two.
The initial masses of black hole and accretion disc and the acccretion radius are rescaled in the same manner.

\begin{figure}
  \includegraphics[width=84mm]{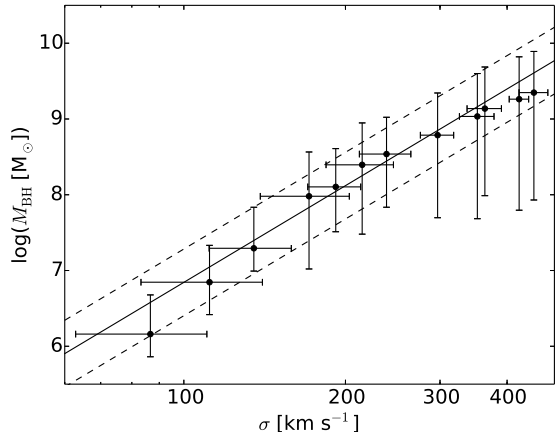}
  \caption{Black hole mass $M_{\rmn{BH}}$ as function of the galaxy's stellar velocity dispersion $\sigma$.
           The dots indicate the black hole mass at the time the black hole accretion rate (BHAR) reaches
           its maximum value. The horizontal bars indicate the error of $\sigma$,
           the vertical bars indicate the range of black hole mass from the time of the end of
           the starburst to the time the BHAR decreases to $0.3\,\%$ of its Eddington rate.
           The solid line indicates the observed $M_{\rmn{BH}}$-$\sigma$ correlation
           with intrinsic scatter (dashed lines) according to \citet{2009_Gueltekin_Richstone_Gebhardt}.}
  \label{fig:mbh_sigma}
\end{figure}

Fig.~\ref{fig:mbh_sigma} shows the mass of the black hole at the time the BHAR reaches its maximum value
as function of the stellar velocity dispersion of the resulting elliptical galaxy,
on how to calculate the latter we refer to \cite{2011_Debuhr_Quataert_Ma}.
The results are consistent with the observed correlation from \citet{2009_Gueltekin_Richstone_Gebhardt} indicated by the solid line.
As the black hole continues growing after the galaxies have merged we additionally plot the range of black hole mass
from the time of the end of the starburst to the time the BHAR decreases to $0.3\,\%$ of its Eddington rate.
According to \citet{2002_Beckert_Duschl} the radiation efficiency of accretion discs declines steeply below this rate,
making it increasingly unlikely to detect those discs.
Our results suggest that this continuing evolution of the black hole mass may contribute to the large scatter of  
the observed $M_{\rmn{BH}}$-$\sigma$ correlation that is indicated by the dashed lines in Fig.~\ref{fig:mbh_sigma}.

\section{Parameter study}\label{sec:param}
As our simulations depend on a number of parameters, some of which are not well constrained by observations,
we will perform a parameter study in this chapter to test the robustness of our results against parameter changes.

However, we only use our reference model with $v_{\rmn{200}}=160\,\rmn{km}\,\rmn{s}^{-1}$ for the parameter study and do not investigate the
effects of parameter changes on the $M_{\rmn{BH}}$-$\sigma$ correlation.
The latter has been done by several authors, e.g.,
\citet{2012_Debuhr_Quataert_Ma,2011_Debuhr_Quataert_Ma,2009_Johansson_Naab_Burkert,2005_Di-Matteo_Springel_Hernquist}.

For our study we choose the following parameters:
the accretion radius $R_{\rmn{acc}}$, the viscosity parameter $\xi$ and the
feedback parameters $\tau_{\rmn{r}}$ and $\tau_{\rmn{w}}$.
The chosen values of these parameters are summarized in Table~\ref{tab:parameterstudy}.
We furthermore investigate the effect of varying the resolution of our simulations.

\begin{figure*}
  \includegraphics[width=177mm]{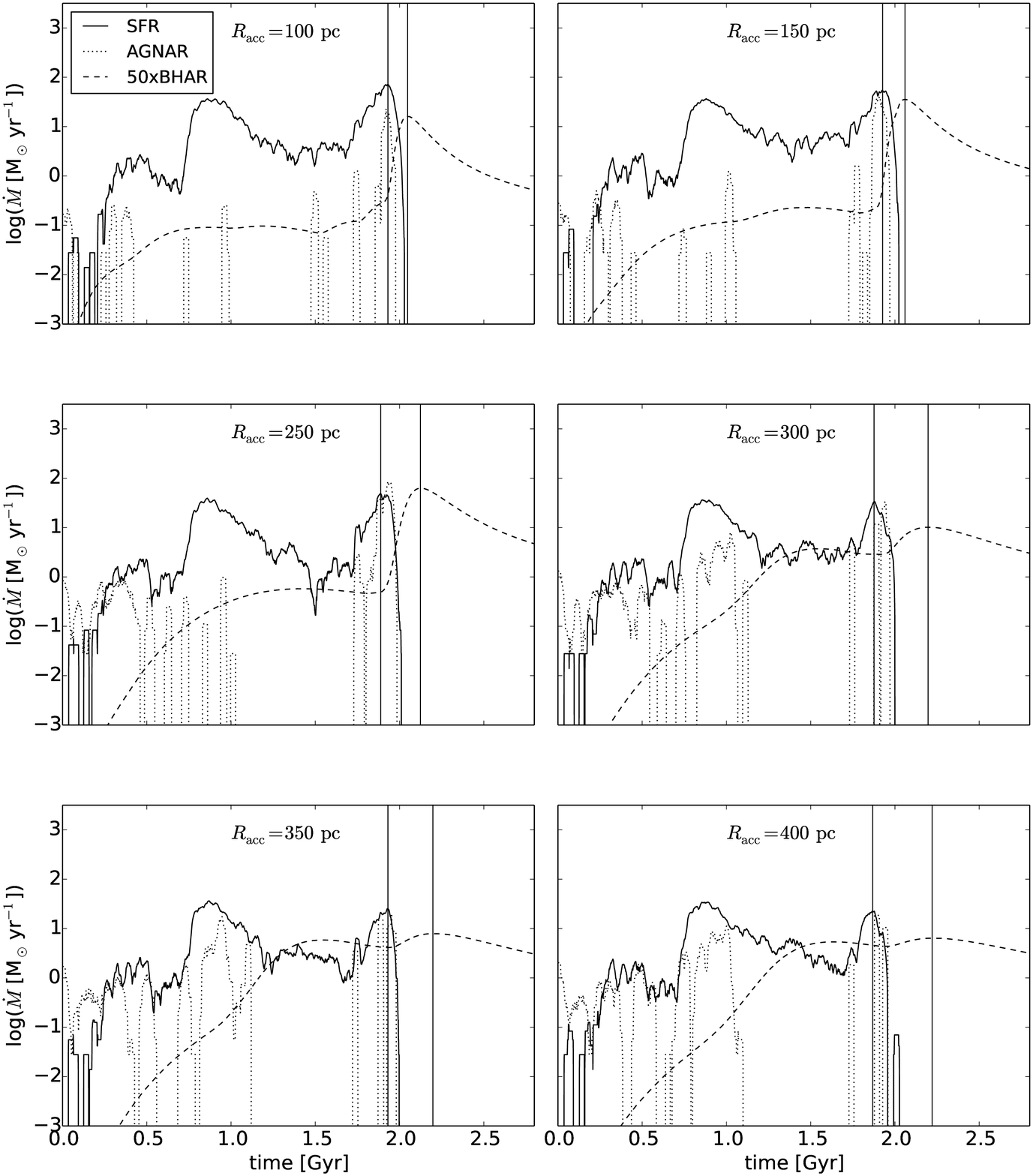}
  \caption{Star formation rate (SFR), accretion rate of the AGN (AGNAR) 
           and black hole accretion rate (BHAR) as functions of time
           for different values of the accretion radius $R_{acc}$.
           The vertical lines mark the peak of the SFR and the peak
           of the BHAR, respectively.}
  \label{fig:massflows_param_Racc}
\end{figure*}

\begin{figure*}
  \includegraphics[width=177mm]{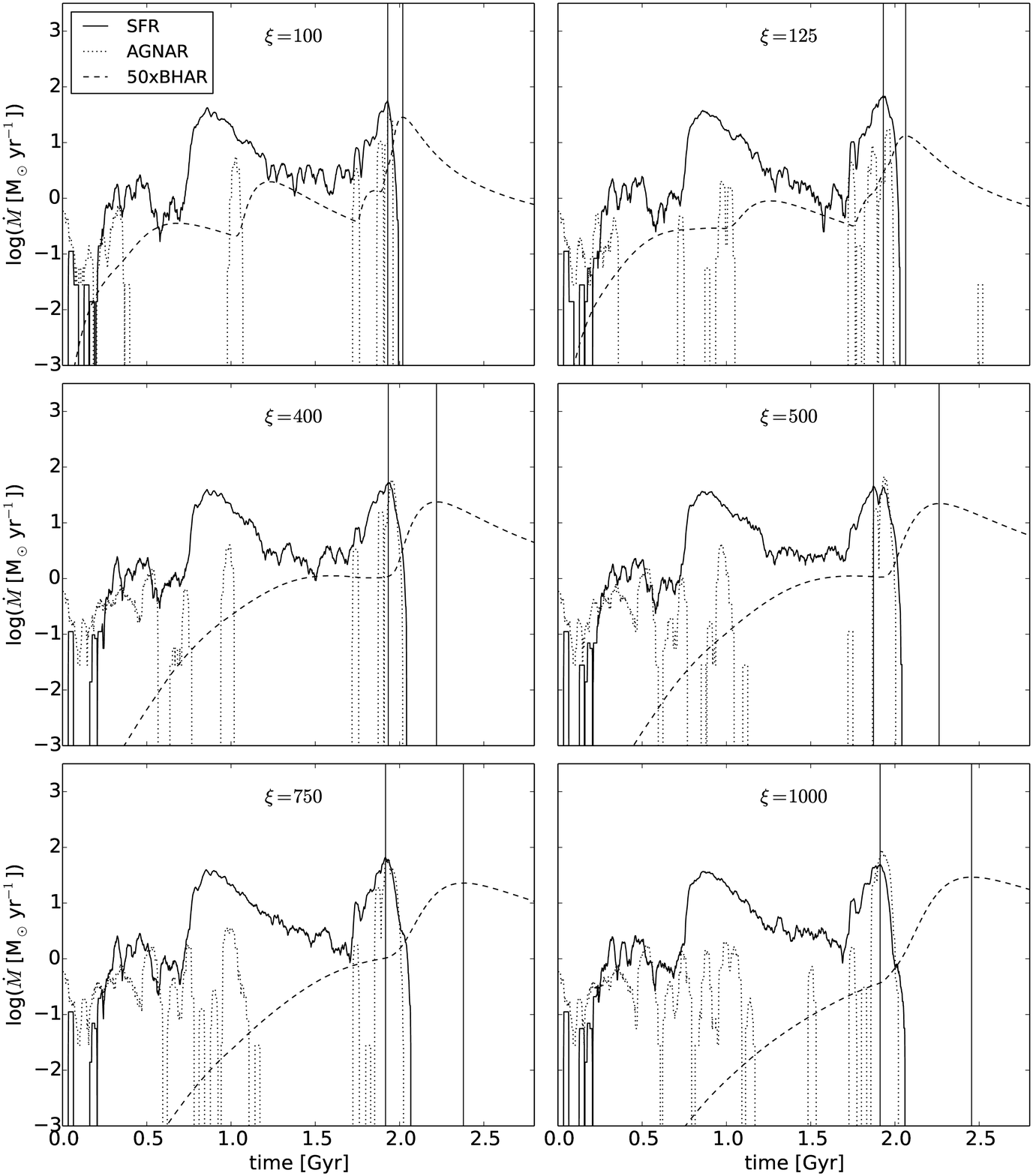}
  \caption{Star formation rate (SFR), accretion rate of the AGN (AGNAR) 
           and black hole accretion rate (BHAR) as functions of time
           for different values of the viscosity parameter $\xi$.
           The vertical lines mark the peak of the SFR and the peak
           of the BHAR, respectively.}
  \label{fig:massflows_param_xi}
\end{figure*}

\begin{figure*}
  \includegraphics[width=177mm]{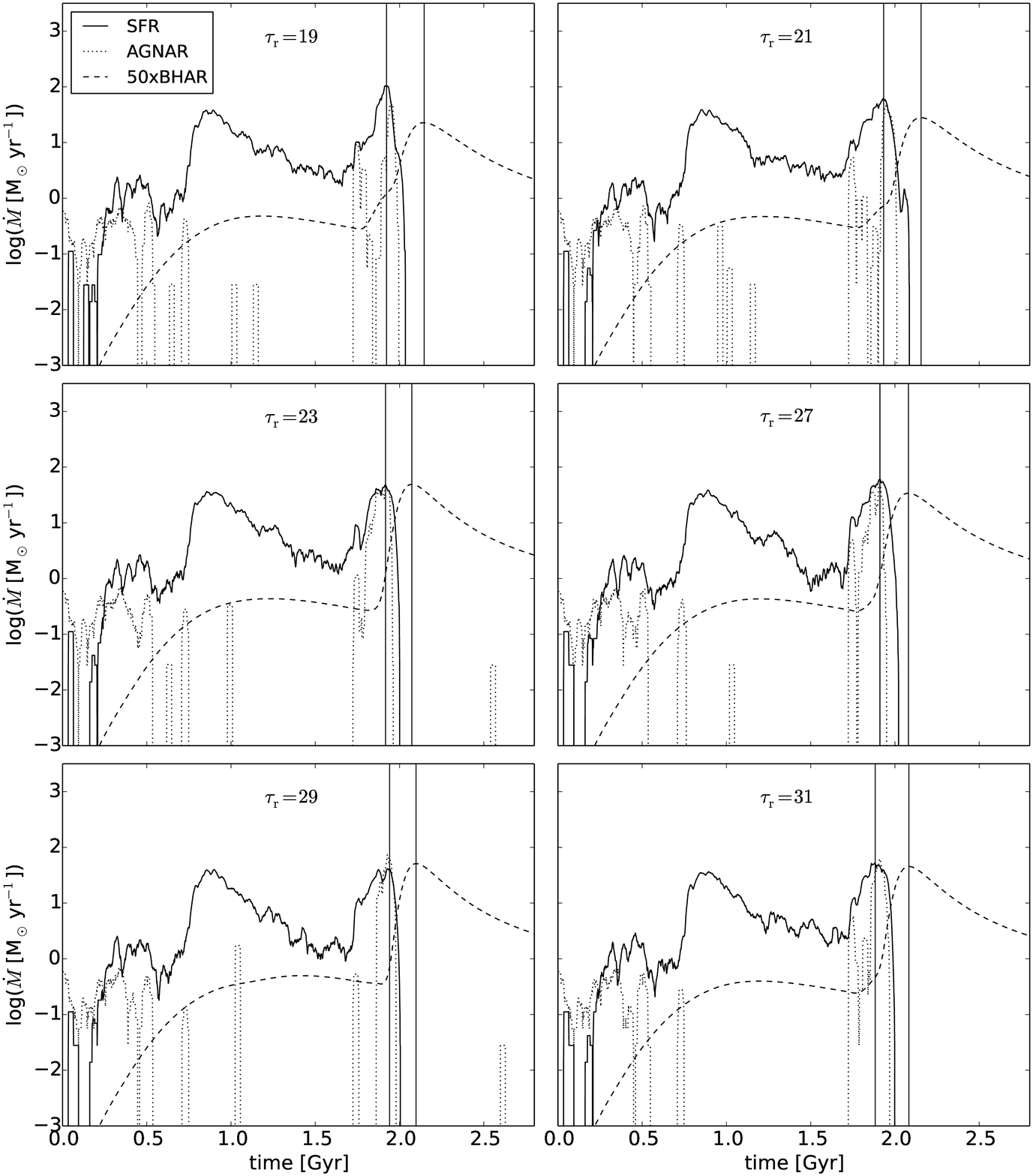}
  \caption{Star formation rate (SFR), accretion rate of the AGN (AGNAR) 
           and black hole accretion rate (BHAR) as functions of time
           for different values of the feedback parameter $\tau_{r}$.
           The vertical lines mark the peak of the SFR and the peak
           of the BHAR, respectively.}
  \label{fig:massflows_param_taur}
\end{figure*}

\begin{figure*}
  \includegraphics[width=177mm]{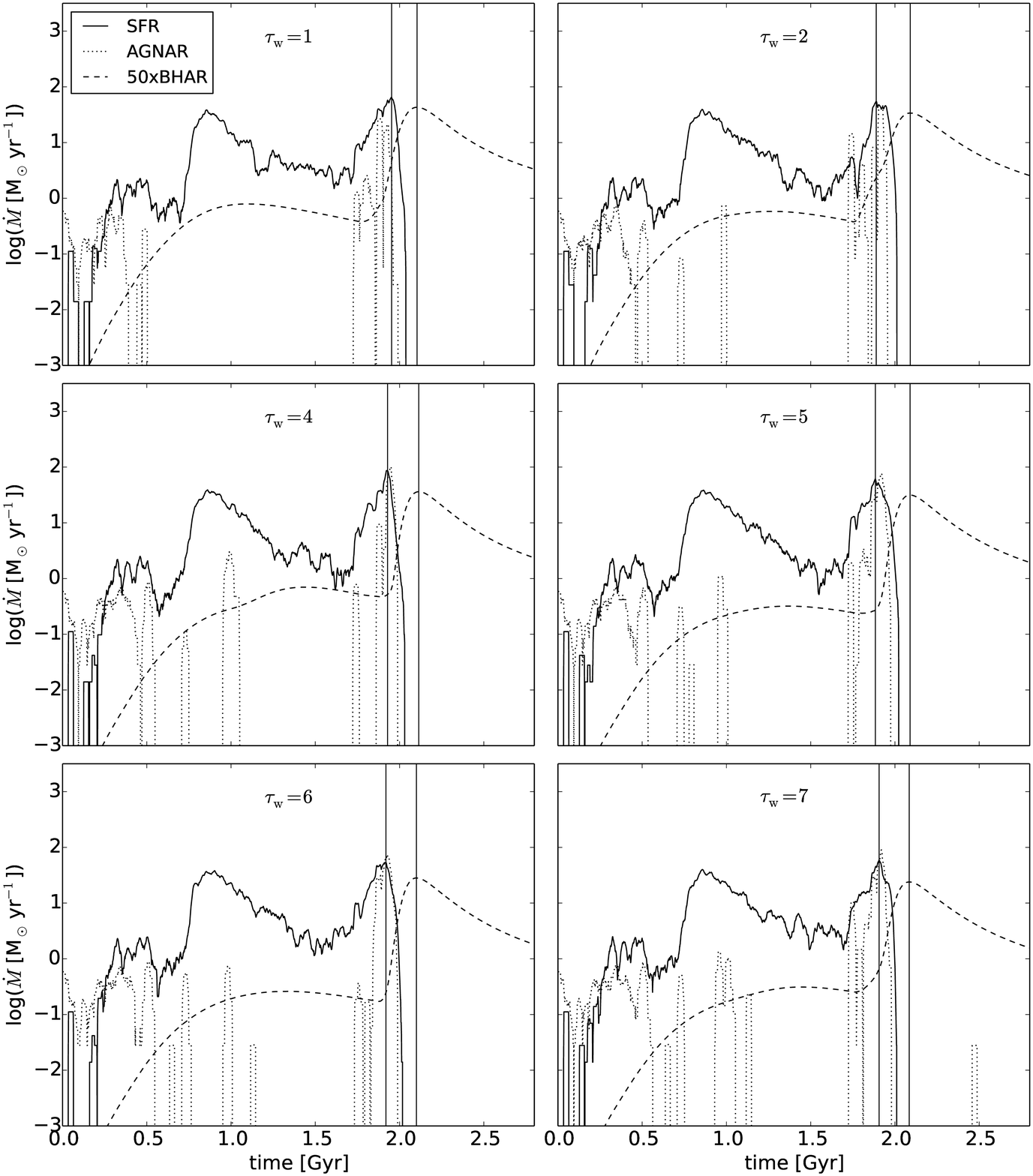}
  \caption{Star formation rate (SFR), accretion rate of the AGN (AGNAR) 
           and black hole accretion rate (BHAR) as functions of time
           for different values of the feedback parameter $\tau_{w}$.
           The vertical lines mark the peak of the SFR and the peak
           of the BHAR, respectively.}
  \label{fig:massflows_param_tauw}
\end{figure*}

We analyze the SFR, BHAR and AGNAR versus time, which are plotted in Fig.~\ref{fig:massflows_param_Racc}
for different accretion radii $R_{\rmn{acc}}$, in
Fig.~\ref{fig:massflows_param_xi} for different viscosity parameters $\xi$
and in Figs. \ref{fig:massflows_param_taur} and \ref{fig:massflows_param_tauw}
for different feedback parameters $\tau_{\rmn{r}}$ and $\tau_{\rmn{w}}$, respectively.
All simulations produce a starburst and a peak in AGN activity as shown for our reference model,
and produce a time lag between these two phenomena of order of some 100\,Myr.
Furthermore all simulations are consistent with the $M_{\rmn{BH}}$-$\sigma$ correlation
in the sense that the black hole mass at the time the BHAR peaks lies within the intrinsic scatter
of the observed correlation of \citet{2009_Gueltekin_Richstone_Gebhardt}.

\begin{table}
  \caption{Parameters used in our parameter study.}
  \label{tab:parameterstudy}
  \begin{tabular}{ll}
    \hline
      $R_{\rmn{acc}}$ (pc)   &   100, 150, 250, 300, 350, 400     \\
      $\xi$                  &   100, 125, 400, 500, 800, 1000    \\
      $\tau_{\rmn{r}}$       &   19, 21, 23, 27, 29, 31           \\
      $\tau_{\rmn{w}}$       &   1, 2, 4, 5, 6, 7                 \\
    \hline
  \end{tabular}
\end{table}

However, the simulations with $R_{\rmn{acc}} = 350\,\rmn{pc}$ and $R_{\rmn{acc}} = 400\,\rmn{pc}$
show only a very weak peak in the BHAR.
In these simulations the first close encounter of both galaxies already drives a huge amount of mass
towards the centre that is accreted by the ADP, as a result there is not much gas left for
accretion when the galaxies finally merge.
According to \citet{2013MNRAS.431..539W} an accretion radius that is too large for the adopted
level of resolution produces a BHAR that is too high, which happens here after the first
encounter of the galaxies.

\begin{figure*}
  \includegraphics[width=177mm]{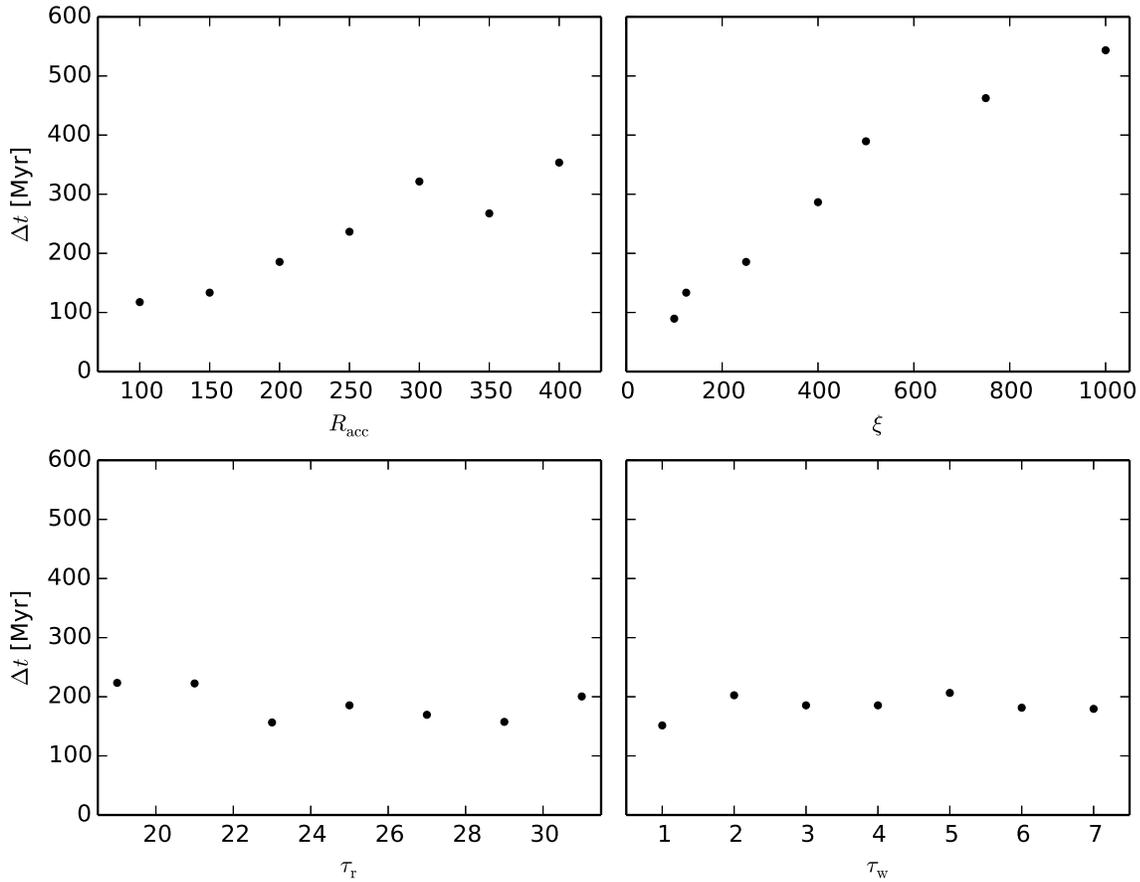}
  \caption{Time lag between starburst and AGN activity with the values of
           $R_{\rmn{acc}}$, $\xi$, $\tau_{\rmn{r}}$ and $\tau_{\rmn{w}}$ from
           our parameterstudy (Table~\ref{tab:parameterstudy}) and our reference model.}
  \label{fig:timing}
\end{figure*}

We also investigate the time lag between starburst and AGN activity.
The time of the starburst we roughly estimate by determining the time the SFR reaches its maximum value,
likewise the time of AGN activity equals the time the BHAR reaches its maximum value.
Both events are represented by vertical lines in Figs.~\ref{fig:massflows_param_Racc}--\ref{fig:massflows_param_tauw}.
The difference of these two times, i.e., the time lag between starburst and AGN activity,
is shown in Fig.~\ref{fig:timing} for all parameters of our parameter study.

As this time lag is a viscous time-scale, it is related to the orbital time-scale $t_{\omega}$ via
\begin{equation}
t_{\nu} = \xi t_{\omega} = 2 \rmn{\pi} \xi \sqrt{\frac{R^3}{GM}}
\end{equation}
where $R$ can be approximated by the accretion radius $R_{\rmn{acc}}$ and $M$ by the sum of black hole mass
and the mass of the accretion disc.
Fig.~\ref{fig:timing} confirms that the time lag increases with increasing $R_{\rmn{acc}}$ and with increasing $\xi$,
and does not change significantly with $\tau_{\rmn{r}}$ and $\tau_{\rmn{w}}$.

Observations show time lags of about 50--250\,Myr which are reproduced by most of our simulations,
but a few of our simulations produce time lags that have values of up to 550\,Myr.
Such large time lags have not been directly observed yet. However, 
\citet{2010MNRAS.405..933W} only measure the average time lag of a sample of about 400 galaxies with a result of
250\,Myr, thus some of the galaxies involved will have time lags that exceed this average value.

\begin{figure*}
  \includegraphics[width=177mm]{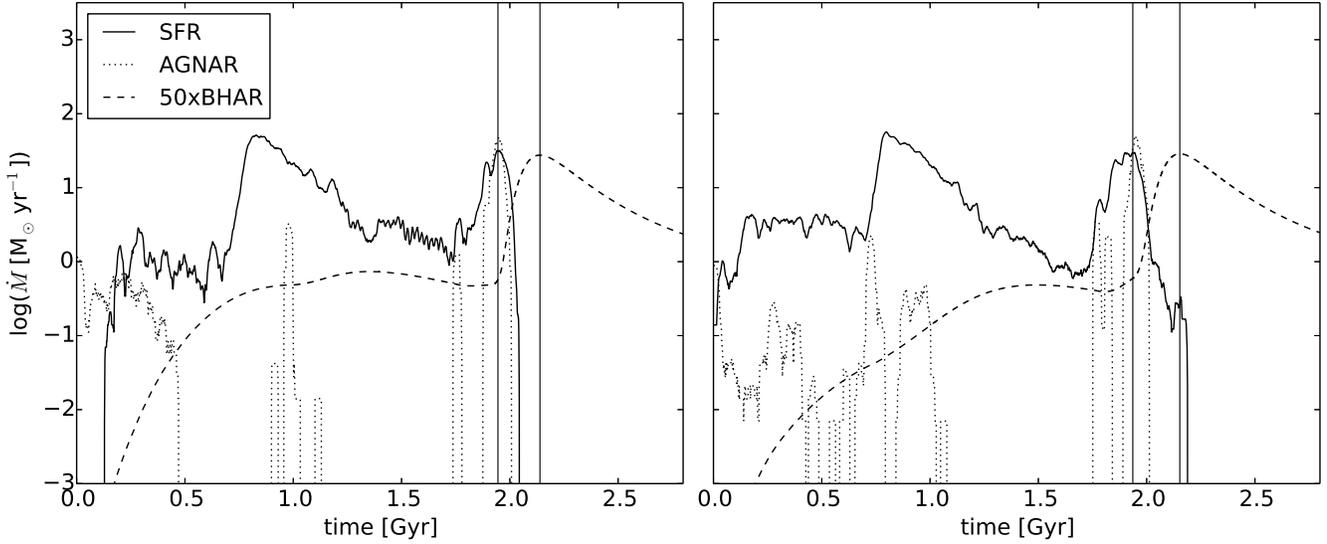}
  \caption{Star formation rate (SFR), accretion rate of the AGN (AGNAR) 
           and black hole accretion rate (BHAR) as functions of time
           for different resolution levels.
           Left: particle number increased by a factor of two, right:
           particle number increased by a factor of four.
           The vertical lines mark the peak of the SFR and the peak
           of the BHAR, respectively.}
  \label{fig:massflows_param_res}
\end{figure*}

Fig.~\ref{fig:massflows_param_res} shows SFR, BHAR nad AGNAR versus time for different levels of resolution.
We increase the number of particles by a factor of two and a factor of four, respectively.

For the former resolution level the gravitational softening length for the halo particles is 0.64 kpc and 0.4 kpc
for all other particles, for the latter resolution level the gravitational softening length for the halo particles
is 0.5 kpc and 0.32 kpc for all other particles.
Both resolution levels show the same features as our reference model: 
the simulations produce a starburst and a peak in the BHAR,
and a time lag between these phenomena of about 200\,Myr.
However, there are still small differences between these simulations, to further reduce these
deviations additional adjustments to the respective resolution levels would be necessary.
For instance, \citet{2013_Hopkins_Narayanan_Murray} argue that the critical density for the onset
of star formation $\rho_{\rmn{crit}}$ has to be adjusted for resolution.
But as the above mentioned key features are already reproduced we refrain
from such a fine tuning of our simulation parameters.

\section{Summary}\label{sec:summary}
With only one model setup we were able to reproduce three observational findings that have been identified in galaxies:

(i) The observed time lag between starburst and AGN activity is, in our work, principally caused by a viscous time
lag the gas needs to flow through the AGN's accretion disc until it reaches the central black hole.

(ii) Our results match the observed $M_{\rmn{BH}}$-$\sigma$ correlation, but additionally include the aforementioned time lags.
As, e.g., \citet{2005_Di-Matteo_Springel_Hernquist} or \citet{2011_Debuhr_Quataert_Ma} have already shown,
AGN feedback is responsible for this relation.

(iii) The large scatter of the $M_{\rmn{BH}}$-$\sigma$ correlation is, in our work, caused by the continuing evolution
of the black hole mass after the merging event.

The time lag does not only occur between starburst and AGN activity, but also between the peak in the AGNAR and
the peak in the BHAR. Thus it is evident that this time lag is caused by the gas being delayed by the AGN particle,
i.e. by the gas needing some time to move through the accretion disc.
In addition our parameter study shows that the time lag increases with increasing size of the accretion disc,
and decreases with increasing viscosity, indicating that it is indeed a viscous time-scale that delays the gas.
Furthermore the parameter study shows that a time lag between starburst and AGN activity of the order of
some 100\,Myr can be reproduced for a wide range of parameters and for different resolution levels.
Thus our conclusions do not depend on the detailed choice of the parameters, as long as they are taken from
within a physically sensible range.





\section*{Acknowledgements}
We thank Drs.\ Ralf Klessen, Volker Springel and Meng Xiang--Gr\"{u}\ss{}
for making available some of the numerical tools that we used for the
calculations presented in this paper.
We furthermore thank the referee for his comments on an earlier
version of this paper, that significantly improved the quality of
the publication.

\bibliographystyle{mn2e}
\bibliography{library}

\bsp

\label{lastpage}

\end{document}